# Lattice location of ion-implanted $^6$He in diamond

U. Wahl,[1,*] J. G. Correia,[1] A. Costa,[1] B. Biesmans,[2] G. Magchiels,[2] S. M. Tunhuma,[2] A. Lamelas,[3] A. Vantomme,[2] L. M. C. Pereira,[2] and the ISOLDE Collaboration[4]

[1] *Centro de Ciências e Tecnologias Nucleares, Departamento de Engenharia e Ciências Nucleares, Instituto Superior Técnico, Universidade de Lisboa, 2695-066 Bobadela LRS, Portugal*

[2] *KU Leuven, Quantum Solid State Physics, 3001 Leuven, Belgium*

[3] *CICECO, Universidade de Aveiro, 3810-193 Aveiro, Portugal*

[4] *CERN-EP, 1211 Geneva 23, Switzerland*

The properties of He in diamond are of interest with respect to geological dating and the creation of color centers with possible quantum applications. We report on the lattice location of the short-lived ion implanted nuclear probe $^6$He ($t_{1/2}$=807 ms) in diamond, which was performed using the beta emission channeling method at CERN's ISOLDE facility. $^6$He was implanted with 30 keV into a single-crystalline chemical-vapor deposited diamond sample kept at a temperature ranging from 30°C up to 800°C. By means of comparing the measured emission channeling patterns along different crystallographic directions with simulated yields for a variety of possible sites, we conclude that the implanted $^6$He occupies tetrahedral (T) interstitial sites, in agreement with theoretical predictions that T sites should be the preferred position of He in diamond. Implantation at 800°C resulted in a drop in the tetrahedral interstitial fraction by ~18%, which we interpret as the onset of diffusion, $^6$He thus being able to change to lattice sites of low crystallographic symmetry, or reach the surface of the sample or escape to the bulk during its lifetime. We estimate the activation energy for interstitial migration of He to be around 1.63-2.89 eV, which agrees with theoretical predictions of 1.41-2.36 eV from the literature. Activation energies around 2 eV would mean that simple interstitial He cannot be stable in diamond on geological time scales, thus to remain inside, it should be bound to some defect in the material or exist in another form such as within inclusions of other minerals or liquids, or possibly small He bubbles.

## I. INTRODUCTION

The ion implantation of He into diamond has been found to create color centers that exhibit luminescence with zero phonon lines (ZPLs) at wavelengths of 536 nm and 560 nm, the so-called "He-related" HR1 and HR2 defects [1-12]. Following implantation of $He^+$ at elevated fluences ($5\times10^{14}$-$5\times10^{16}$ atoms/cm$^2$), the ZPLs appeared for annealing temperatures around 550°C for 1 h [10], 600°C for 1 h [1,11], 625°C [12] or 750°C for 2 h [7] and their intensity was found to be maximized after annealing around 850°C [2], 950°C [10], 1050°C [3,12], and 1100°C [11], and correlated with the implanted He fluence [6-8,11]. The attribution of the ZPLs to He rather than implantation-related defects was suggested due to the fact that implantations of other light elements such as H [3], C [3,7], N [1], or Ne [2] did not seem to create the HR1 or HR2 defects, although there is a single reference [1] which also reported the same ZPLs following the implantation of Ne. Various tentative models of these centers have been suggested, e.g., "neutral single He atoms disturbed by various radiation defects" [4], neutral He-vacancy defects [5], "complexes containing both He impurities and (most likely) vacancies" [7], centers that "lack inversion symmetry" [8], and where "interstitials are involved" [12], and "a defect with either a $C_2$ or $C_{2v}$ symmetry, most likely $He_S^0$" (neutrally charged substitutional He) [11].

He ion irradiation is also being used in another technological application that involves processing of diamond for the creation of single photon emitting color centers, in this case to create vacancies that can form nitrogen-vacancy (N*V*) centers within a well-defined specific depth interval near the surface of the sample [9,13-18]. Furthermore, high-fluence ($6.4\times10^{16}$ atoms/cm$^2$) high-temperature (1300°C) implantation of 275 keV He in diamond resulted in the formation of small platelets that were identified as solid He under pressure as high as 166 GPa inserted into (100) diamond planes [19].

Apart from actively modifying the properties of diamond, significant scientific interest in the behavior of He is also due to the hypothesis that the amount of $^4$He and the $^3$He/$^4$He ratio found within the material and its inclusions can be used to date terrestrial diamonds [20-24] or learn about the origins of meteoritic nano-diamonds [25-28].

Among the relevant issues in all of these scientific topics are the lattice location of He as well as its diffusion properties. Even under geological aspects, this also applies to the ion-implanted state, since part of $^4$He is introduced into the material due to the alpha decay from the U and Th decay chains [20-24,28,29], while $^3$He is mainly of primordial origin. With respect to $^3$He/$^4$He based dating, the diffusion behavior of He is of particular relevance since possible out-diffusion of any He species at elevated temperatures (typically around 900-1400°C in the earth mantle) and on geological time scales ($10^9$ years) could alter the outcome of such analyses.

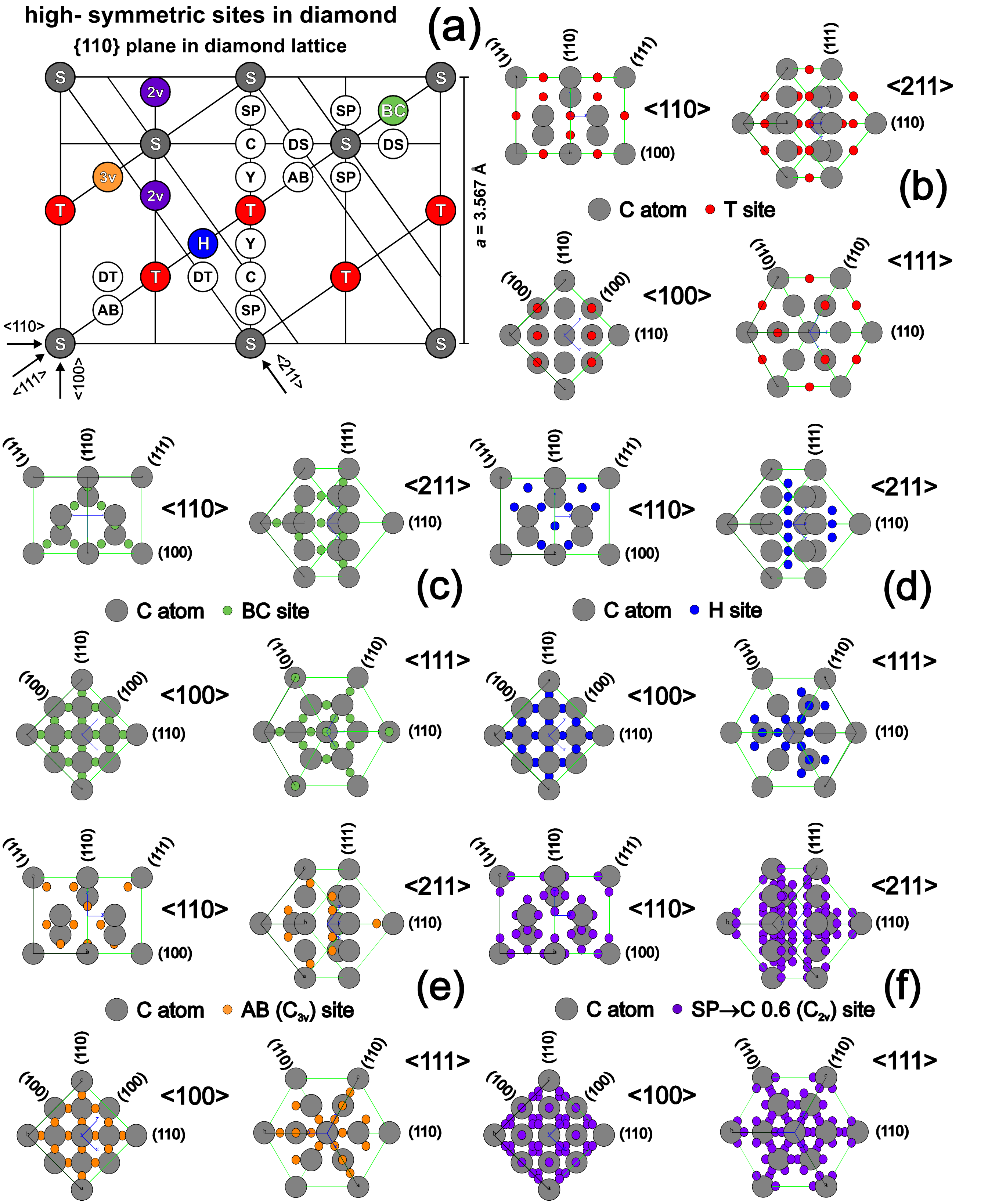

FIG 1. (a) Most common sites of high crystal symmetry in the diamond lattice, which are the substitutional site (S) and a variety of interstitial sites: tetrahedral (T), hexagonal (H), bond-center (BC), anti-bonding (AB), <100>-split (SP), <110>-split (DS), as well as so-called Y, C and DT sites. "3v" and "2v" indicate positions of $C_{3v}$ and $C_{2v}$ symmetry that were proposed for He in Ref. [35] following its introduction as a substitutional atom and relaxation; they are basically identical to AB sites and positions which are displaced from S sites by 0.7 Å along <100> directions, i.e., around 0.6 times the distance from SP to C sites (SP→C 0.6), respectively. (b)-(f) Schematic atomic projections along the major crystallographic directions of the diamond unit cell. The large grey circles show the C atom substitutional sites S. The positions of T sites within the unit cell are shown as small red circles in panel (b), BC sites as small green circles in (c), H sites as small blue circles in (d). The small light orange circles in panel (e) show the AB sites ($C_{3v}$ symmetry) and the small violet circles in panel (f) the (SP→C 0.6) positions ($C_{2v}$ symmetry). Note that projections for T sites are identical to those of S sites along <100> and <111> (six outer atomic rows are not shown, lying outside the unit cell), and those of BC, H and AB sites are identical along <100> and <111>.

TABLE I. Suggested He lattice sites, as well as migration energies $E_M$ or activation energies $E_A$ and entropy constants $D_0$ for He diffusion in diamond reported in the quoted references. A cell in the table being empty, means no information was given on that item, e.g., none of the theoretical works explicitly gives a value for $D_0$, while it seems reasonable it should be calculated according to eq. (2), as we did. The remarks mean the following. $^{10}B(n,\alpha)^7Li$: $^4He$ was introduced into natural $^{10}B$-containing diamond via neutron irradiation; MS: He release was measured by Mass Spectrometry during annealing; natural $^4He/^3He$: terrestrial diamonds containing natural amounts of $^4He$ and $^3He$; NRA: $^3He$ was probed by $^3He(d,p)^4He$ Nuclear Reaction Analysis.

| Ref. | Suggested He lattice sites | $D_0$ [$cm^2s^{-1}$] | $E_M$ or $E_A$ [eV] | Remarks |
|---|---|---|---|---|
| [30] | T, 0.3 Å from S, split-vacancy | | | theory, tight-binding potential |
| [31] | T | | 2.3 | theory, DFT |
| [32] | T, multi-vacancy complexes | | 1.41 | theory, atomistic ab-initio and tight-binding |
| [33] | T | | 1.97 | theory, DFT |
| [34] | T, S→AB 0.65, split-vacancy | | | theory, DFT |
| [35] | T, AB, SP→C 0.60 | | 2.36 | theory, DFT |
| [38] | possibly 30% T + 70% random | | | implanted $^3He$ in natural diamond, NRA |
| this work | T | ~$5\times10^{-3}$ | 1.63-2.89 | $D_0$ assumed from theory, $E_A$ estimated from experiment |
| [39] | | $7\times10^{-4}$ | 1.01 | $^4He$ introduced in natural diamond via $^{10}B(n,\alpha)^7Li$, MS |
| [40] | | $1.8\times10^{21}$ | 11.7 | natural $^4He$, MS |
| [29] | | $6\times10^{-11}$ | 1.55 | natural $^3He$, MS |
| [41] | | $1.6\times10^{-13}$ | 1.07 | natural $^4He$, MS |
| [33] | | $4\times10^{-11}$ | 1.43 | implanted $^3He$ in CVD diamond, NRA |

Hence, quite some theoretical works [30-35] have addressed He in diamond and made predictions for its most stable sites (see Fig. 1), their energies of formation, as well as for the activation energy of interstitial He migration. Table 1 includes a brief overview of these theoretical studies, while some further details are to be found in the Supplemental Material [36]. Almost all studies proposed tetrahedral interstitial T sites as the most stable lattice site of He in diamond. Substitutional S sites were considered, but they were found to have higher energies of formation than $He_T$ and undergo displacements of about 0.3 Å [30] and 0.5 Å [34], with Ref. [35] providing detailed geometries: it was predicted that He on substitutional sites should undergo large lattice relaxations that move it a considerable distance away from the ideal substitutional position, 0.695 Å along the <100> direction ($C_{2v}$ symmetry), which would be the SP→C 0.60 site with our nomenclature, see Fig. 1 (f), or 0.794 Å along the <111> direction ($C_{3v}$ symmetry) which is very close to the AB site, see Fig. 1 (e).

A single He atom inside a double vacancy was also considered in some cases, and found to have the so-called split-vacancy configuration, where the He occupies the same site as a bond-centered (BC) position in the perfect diamond lattice, but in this case with two adjacent vacancies [31,34,35]. Ref. [30] also addressed the structure and stability of possible small He inclusions in diamond, i.e., several He atoms inside multi-vacancy complexes. It was found that "a single carbon vacancy can accommodate up to 4 helium atoms in an energetically favourable position compared to the interstitial defect", and with respect to multiple vacancies that "the most stable site of helium in diamond is in the cavities".

With respect to He migration, the studies agreed that hexagonal (H) interstitial sites would be the saddle points of the atomic jumps, hence a T-H-T migration path [31-33,35]. The predicted activation energies $E_M$ for interstitial diffusion (Table 1) were in the range 1.41-2.36 eV.

Experimental results on the lattice location of He in semiconductors are scarce but there exists some limited work on diamond and its close relatives Si and 6H-SiC [37,38]. Ion beam channeling studies following high-fluence implantation (4-$5\times10^{15}$ $cm^{-2}$) of $^3He$ in natural diamond and Si [37], as well as 6H-SiC [38], indicated, besides large fractions of He on random sites, minor amounts of He on interstitial sites, which were most likely T sites in all three cases. The published data in [37] allow us to estimate that probably only ~33% of He in diamond occupied T sites in such high-fluence experiments. Some more detailed comments on these studies are to be found in the Supplemental Material [36].

There are quite some discrepancies in the literature regarding the activation energies $E_A$ and entropy constants $D_0$ for He diffusion in diamond reported from experiments (Table 1). Note that we use the term $E_A$ instead of $E_M$ for the experimental results because likely the underlying diffusion processes are not due to simple interstitial migration. While most studies reported activation energies around 1.0–1.6 eV, which would be roughly in agreement with the theoretical predictions for simple interstitial migration ranging from 1.41 eV to 2.36 eV, the reported entropy constants were in most cases substantially lower than the expected $D_0$ according to eq. (2), which is around $D_0$~$5\times10^{-3}$ $cm^2s^{-1}$. This strongly suggests that it is not simple interstitial He migration that was the dominant diffusion mechanism observed during these experimental studies. It has been addressed previously in the literature that the diffusion behavior of He in diamond suggests the existence of several different diffusion and release mechanisms, that critically depend on the structural form how He is incorporated in the material, which, e.g., also quite likely differs between $^4He$ resulting from α-decay and cosmogenic $^3He$ [29,33]. For instance, the diffusion data extracted from Ref. [39] are similar to the ones expected for simple interstitial migration. Remarkably, in that study $^4He$ was introduced into B-doped diamond via the nuclear reaction $^{10}B(n,\alpha)^7Li$, meaning it was recoil-implanted with an energy around 1.5 MeV and distributed throughout the whole micro-crystalline diamond samples. The out-diffusion of $^4$He could then be observed at temperatures above 1200°C via mass-spectroscopy. In Ref. [33] $^3$He was implanted into artificial diamond, and the amount that remained during thermal annealing above 950°C probed with $^3He(d,p)^4He$ Nuclear Reaction Analysis (NRA). While the activation energy of diffusion was well in accordance with interstitial migration, the derived entropy constant was rather small. The three other Refs. [29,40,41] all concern release of natural $^4$He or $^3$He from terrestrial diamonds. In that case, its release at high temperatures will not be governed merely by interstitial migration since at least some of the $^4$He is contained in inclusions of the material [22-24,29]. The significantly different values for $D_0$ and $E_A$ in Ref. [40] have previously been pointed out in the literature [33], and it was speculated that a possible reason might be that they were only based on data from two different temperatures.

In this work, we provide a detailed and quantitative analysis of the lattice sites of low-fluence implanted He in diamond. By measuring from room temperature up to 800°C with intermediate steps at 300°C and 600°C, we probed for possible thermally activated diffusion processes. We show that the large majority of implanted He is found on tetrahedral interstitial T sites in accordance with theoretical predictions that this should be the most stable He site in diamond [31,33-35]. There was no indication for other highly symmetric He lattice sites in our work and we estimate them to be below 10%. The activation energy $E_M$ for interstitial migration of He in diamond suggested from our study is 1.63-2.89 eV, which is in reasonable agreement with theoretical predictions of $E_M$ to be around 2 eV [31-33,35].

## II. METHODS

For determining the lattice location of implanted He, we used the $\beta^-$ emission channeling (EC) technique from radioactive isotopes [42-46]. In this experimental approach, radioactive probe isotopes are ion implanted into single-crystalline samples, where they occupy specific lattice sites, on which they decay by the emission of $\beta^-$ particles. These fast electrons interact with the crystal potential which guides them on their way out of the sample. This effect results in a pronounced angular dependence of the $\beta^-$ emission yields around major crystallographic directions, which is characteristic for the lattice site(s) that the probe atoms occupied during their decay. The angular-dependent emission patterns are recorded by means of a two-dimensional position-sensitive detector (PSD) [44,45] which is placed at a suitable distance from the sample in order to achieve sufficient angular resolution. The major lattice sites can be identified by least-square fitting the experimentally observed angle-dependent emission yields by linear combinations of theoretically expected patterns [44,45,47], that have been calculated for specific positions of the emitter atoms in the lattice.

From the radioactive isotopes of He, only $^6$He has a sufficiently long half life ($t_{1/2}$ = 806.7 ms) to be delivered in amounts that are suitable for emission channeling lattice location experiments. $^6$He was produced at the online isotope separator facility ISOLDE [48] at CERN, by means of 1.4 GeV proton-induced spallation reactions from a $UC_2$ target. Following out-diffusion from the $UC_2$ target at a temperature around 2000°C, the $^6$He atoms were ionized to the 1+ charge state, electrostatically accelerated to 30 keV and magnetically mass separated. For ionization we used an Ar plasma source optimized for the extraction of noble gases, with a cold (water-cooled Cu) transfer line (VADIS VD7 Versatile Arc Discharge Ion Source [49]). While a hot transfer line VD5 plasma source would have resulted in higher yields of $^6$He [50], tests showed it was not an option due to excessive beam contamination with doubly-charged stable $^{12}C^{2+}$ ions. The EC measurements were performed simultaneously with implantations into a 1 mm diameter beam spot using the online setup described in Ref. [51].

The radioactive isotope $^6$He decays by the emission of $\beta^-$ particles with an endpoint energy of 3508 keV (average $\beta^-$ energy 1569 keV) into stable $^6$Li, see Fig. S1 in the Supplemental Material [36] for the $\beta^-$ spectrum. Since the angular width of channeling effects scales approximately inversely proportional to the square root of the particle energy, $\beta^-$ EC effects from $^6$He have a rather narrow angular width (their critical angles for channeling in diamond will be around 0.40°, i.e., angular widths around 0.8°), requiring good angular resolution. The sample was analyzed using a $3 \times 3$ cm$^2$ Si pad PSD consisting of $22 \times 22$ pixels of size $1.3 \times 1.3$ mm$^2$ [45], which was placed 60 cm from the sample, resulting in an angular range of ±1.36°. While the comparatively large distance from sample to detector guarantees sufficient angular resolution (0.05° standard deviation), it resulted in a relatively small solid angle of detection.

The sample was a <100> oriented single crystal from ElementSix grown by chemical vapor deposition (CVD), termed “SC plate CVD”, of size 3.0×3.0×0.25 mm$^3$, with nitrogen concentration [N]<1 ppm. $^6$He was implanted through a 1 mm diameter collimator with a current of 0.9 pA, corresponding to $5\times10^6$ atoms/s or $6.4\times10^8$ atoms/s/cm$^2$. Emission channeling patterns were typically acquired with a statistics of $(1\text{-}3)\times10^6$ events, see Supplemental Material [36], corresponding to an implanted fluence of ~(0.7-2)×$10^{12}$ atoms/cm$^2$/pattern and $2.9\times10^{13}$ atoms/cm$^2$ at the end of the experiment. According to Stopping and Range of Ions in Matter (SRIM2010, [52]) simulations, the projected range of 30 keV implantation of $^6$He into diamond is around 1375 Å with a straggling of 299 Å and a peak concentration of $1.5\times10^5$ atoms cm$^2$/cm$^3$. We note that due to its short 807-ms half life there is no persistent accumulation of $^6$He in the sample, with its max. concentration remaining at $6.4\times10^8$ atoms/s/cm$^2$ × $1.5\times10^5$ atoms cm$^2$/cm$^3$ × $t_{1/2}$/ln(2) = $1.1\times10^{14}$ atoms/cm$^3$, meaning the average distance between $^6$He atoms remains larger than 2100 Å. The temperature during implantation was varied in between room temperature (RT) of 30°C and 800°C, in vacuum better than $10^{-5}$ mbar.

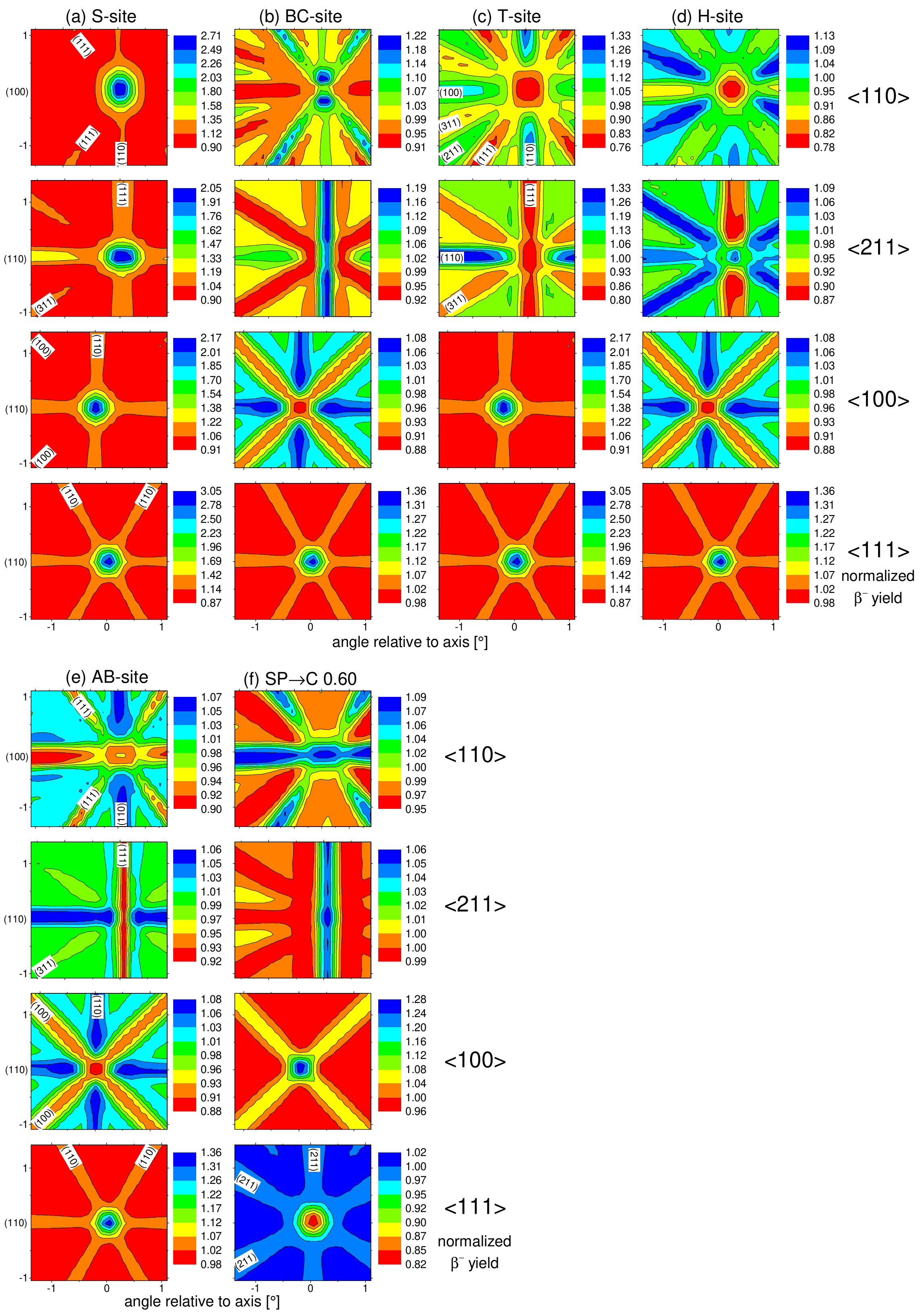

FIG 2. Theoretical angular distributions of β⁻ emission yields around the major crystallographic directions of diamond for $^6$He emitter atoms on (a) S, (b) BC, (c) T, (d) H, (e) AB, and (f) SP→C 0.60 sites. All patterns are displayed so as to reflect the angular range (around ±1.3° in *x* and *y* directions) and orientation of the experimental patterns in Figs. 3-4, taking into account the angular resolution of the experimental data and the size and orientation of the detector pixels. Note that the <111> and <100> patterns from T sites are completely identical to the one from S sites, since T sites are perfectly aligned with the atomic rows along these two crystallographic directions. <111> patterns from BC, H and AB sites are also identical, but while their contrast appears very similar to the <111> patterns from S and T sites, the maximum normalized yield is pronouncedly different: 3.05 for S or T sites vs 1.36 for BC or H sites.

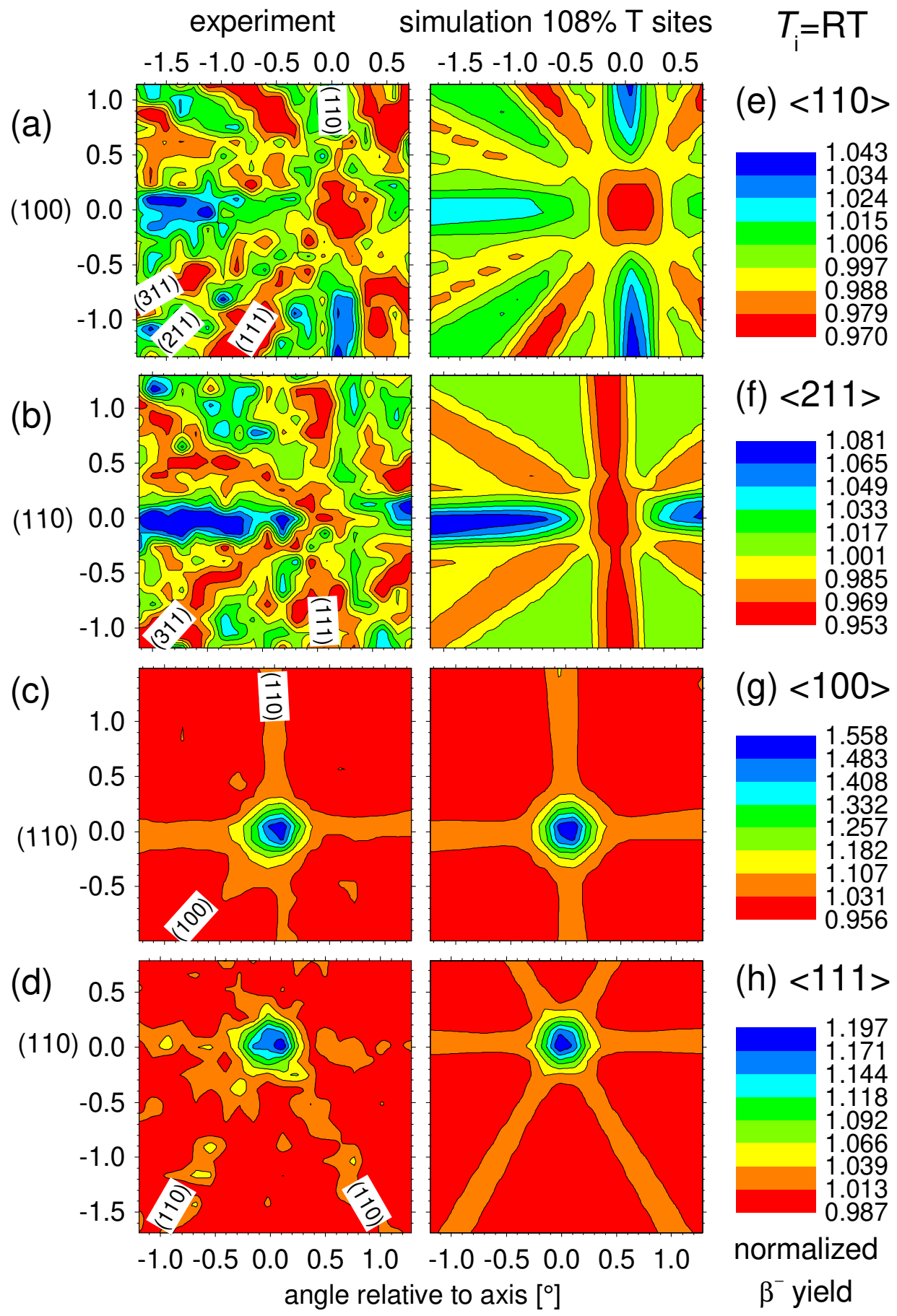

FIG 3. (a)-(d) Angular distributions of $\beta^-$ emission yields from $^6$He in diamond, during room temperature (RT, 30°C) implantation, around the crystallographic directions <110>, <211>, <100>, and <111>. (e)-(h) are the best fits of simulated patterns, corresponding to 108% of $^6$He on tetrahedral interstitial (T) sites.

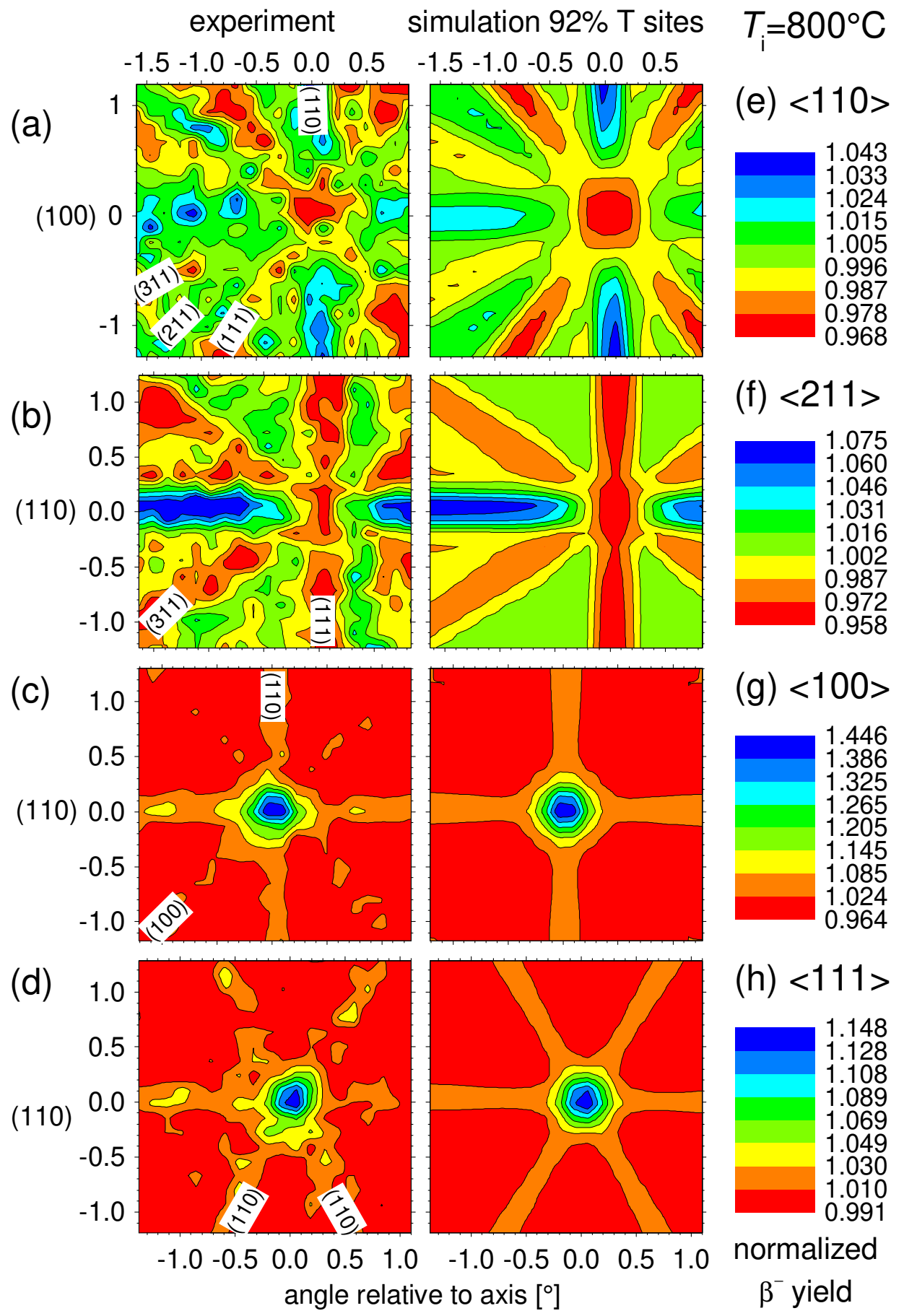

FIG 4. (a)-(d) Angular distributions of $\beta^-$ emission yields from $^6$He in diamond, during implantation at 800°C, around the crystallographic directions <110>, <211>, <100>, and <111>. (e)-(h) are the best fits of simulated patterns, corresponding to 92% of $^6$He on tetrahedral interstitial (T) sites.

When measuring electron emission channeling patterns the recorded angular anisotropy is always subject to background, which arises mostly from two sources: a) electrons which are backscattered from the sample itself, the sample holder, or from the walls of the vacuum chamber and hence do not contribute to the channeling effect, and b) gamma radiation. Both backscattered electron and gamma background were taken into account in the analysis, by multiplying fitted fractions with a background correction factor of 2.36 (details on derivation of the correction factor can be found in Supplemental Material [36]). The procedure to derive the background correction factor is typically subject to absolute errors of ±(10-15)%, which means that after performing background correction it is possible to arrive at total fractions above 100%.

Theoretical emission patterns for a variety of ~250 lattice sites were simulated using the "many beam" approach [42,44], and more details on its specific application to diamond can be found in the Supplemental Material [36] and Refs. [53,54]. We considered lattice sites which were obtained by displacing from S or T sites along <111>, <100>, or <110> directions within the whole diamond unit cell, while keeping a rms thermal displacement of $u_1(^6\text{He})=0.044$ Å.

## III. RESULTS AND DISCUSSION

Figure 1 panels (b)-(f) show the projections along the four major crystallographic directions of the diamond structure for five selected interstitial sites, while panel (a) indicates the positions of these and other major sites within the (110) plane. These five interstitial sites, in addition to the substitutional site S, might be expected as likely candidates for He sites in diamond: (b) the tetrahedral interstitial T site, (c) the so-called bond-center (BC) site which would be expected for He*V* complexes in split-vacancy configuration, (d) the hexagonal interstitial site H, (e) the anti-bonding site AB [35] and (f) the SP→C 0.60 site [35]. Figure 2 shows the theoretically calculated electron emission patterns along all four crystallographic directions studied in this work, for the substitutional sites S plus for the five interstitial positions shown in Fig. 1 (b)-(f).

Panels (a)-(d) of Figs. 3 and 4 display the experimentally found normalized angular distributions of beta particles around all four major directions of diamond [55], measured at implantation temperatures of 30°C (Fig. 3) and 800°C (Fig. 4), the ones for 600°C can be found in Fig. S2 of the Supplemental Material [36]. From a first visual inspection of the experimental patterns by means of comparing to the theoretically expected patterns for S, BC, T, H, AB and SP→C 0.60 sites, as shown in Fig. 2, one arrives at the following conclusions: while the channeling effects observed for <111> patterns would be qualitatively in accordance with the S, BC, T, H, AB sites, they are incompatible with the blocking effect that is expected for the SP→C 0.60 site. Moreover, the channeling effects along <100> are only compatible with S or T sites. Finally, the <110> and <211> patterns clearly allow distinguishing between substitutional S and interstitial T sites, with only interstitial T sites displaying the characteristic combination of axial and planar channeling and blocking features in accordance with the experimental data.

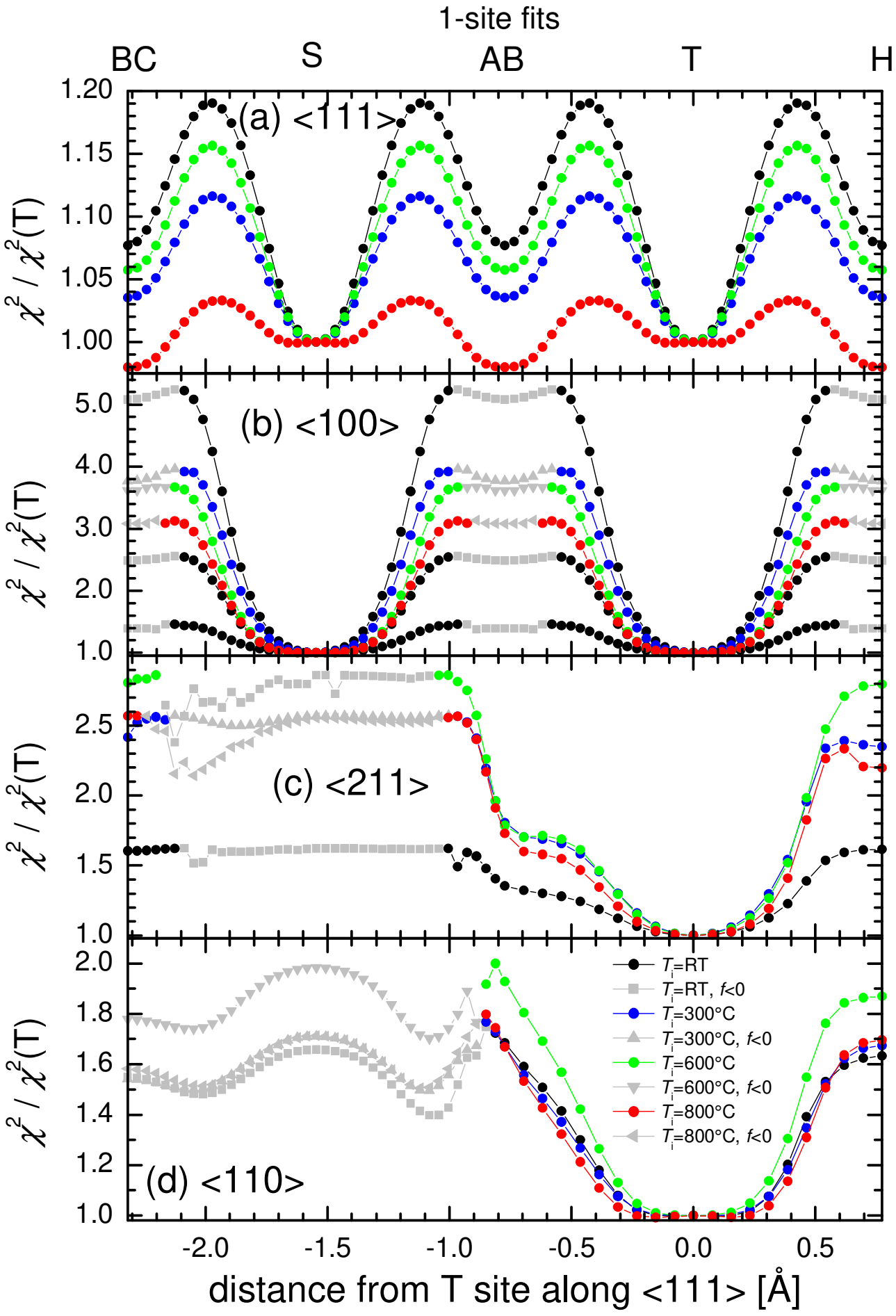

FIG 5. Relative chi square of one-site fits as a function of the $^6$He position along the <111> direction with respect to T sites, given as $\chi^2/\chi^2$(T), i.e., obtained by dividing by the chi square $\chi^2$(T) resulting from $^6$He on T sites. Note that <111> and <100> results in panels (a) and (b) cannot be used to distinguish between all possible sites since, due to the crystal symmetry, their channeling patterns for S and T, or BC, AB and H sites, are equal. The black <100> curves represent three measurements obtained at room termperature but with slightly different orientations towards the detector. The curves shown in light gray color correspond to fits with a negative site fraction ($f$<0), hence physically irrelevant results.

These qualitative considerations were then fully confirmed by means of quantitative fit procedures, where first one-site fits were performed for all ~250 lattice sites considered, which resulted in best fits, i.e., lowest chi square, for T sites. Figure 5 shows the relative chi square of the fits of experimental patterns for all major crystallographic directions investigated, as a function of $^6$He site when its position is varied along <111> directions, starting from BC sites and including S, AB, T and H sites. In case of the <111> and <100> patterns, clear minima of the chi square are found at T or S sites, while the <110> and <211> patterns only show strong global minima at the interstitial T sites. Thus, while channeling patterns of the directions <111> and <100> cannot distinguish between T or S sites due to the crystal symmetry, in combination with the <110> and <211> results it is clear that T sites represent by far the best match for all sites that were tried in these one-site fits. Similarly, we subsequently performed fits where the position of $^6$He was varied along <100> or <110> directions, starting at the T sites. The results, which are shown in Fig. S3 of the Supplemental Material [36], further confirmed clear minima of chi square at the T sites. Note that the shape of the minima does not allow for displacements much larger than 0.1 Å from the ideal T sites.

In a next step, we performed two-site fits, where the first site was kept fixed at the T position, while the second site was allowed to vary among all the ~250 sites that had been simulated. The results are shown in Fig. S4 of the Supplemental Material [36]. While there are in some cases relative improvements in chi square up to 10%, these are quite often not consistent among the four different crystallographic directions investigated or associated with at least one of the two fitted site fractions becoming negative, i.e., unphysical results. Chi square improvements of 2-8% for <110> and <211> pattern fits are obtained for fractions of 2-10% on S sites, so an occupation of up to 10% on S sites might be a possibility (we already noted that in case of <100> and <111> patterns, S and T sites cannot be distinguished). On the other hand, the He sites that were theoretically predicted to result from a relaxation of substitutional He, i.e., the SP→C 0.60 or AB sites [35], do not result in consistent chi square improvements, often being associated also with negative fractions. With respect to the He defects that act as color centers [1-12], these were always obtained for rather high fluences of implanted He ($5\times10^{14}$-$5\times10^{16}$ atoms/cm$^2$, i.e., 10-1000 times higher than in our study). It thus seems possible that they have such low production yields, e.g., a few per cent only, that they are not detectable in emission channeling experiments. Summarizing, only T sites can be clearly identified as major lattice sites occupied by implanted He, although it cannot be excluded that minor fractions of $^6$He (our estimate is up to 10%) occupy other sites.

Panels (e)-(h) in Figs. 3, S2 and 4 show the best one-site fits of theoretical patterns to the experimental data, which were obtained for interstitial T sites and fractions of 108% at RT=30°C, 111% at 600°C, and 92% at 800°C (note that our experimental setup does not allow us to perform measurements in the temperature range above 800°C, so that $^6$He lattice sites could only be probed up to that temperature). As mentioned above in the Methods section, fractions above 100% may result from an overestimation of the background correction, see also the comments regarding the precision vs accuracy of derived fractions in the Supplemental Material

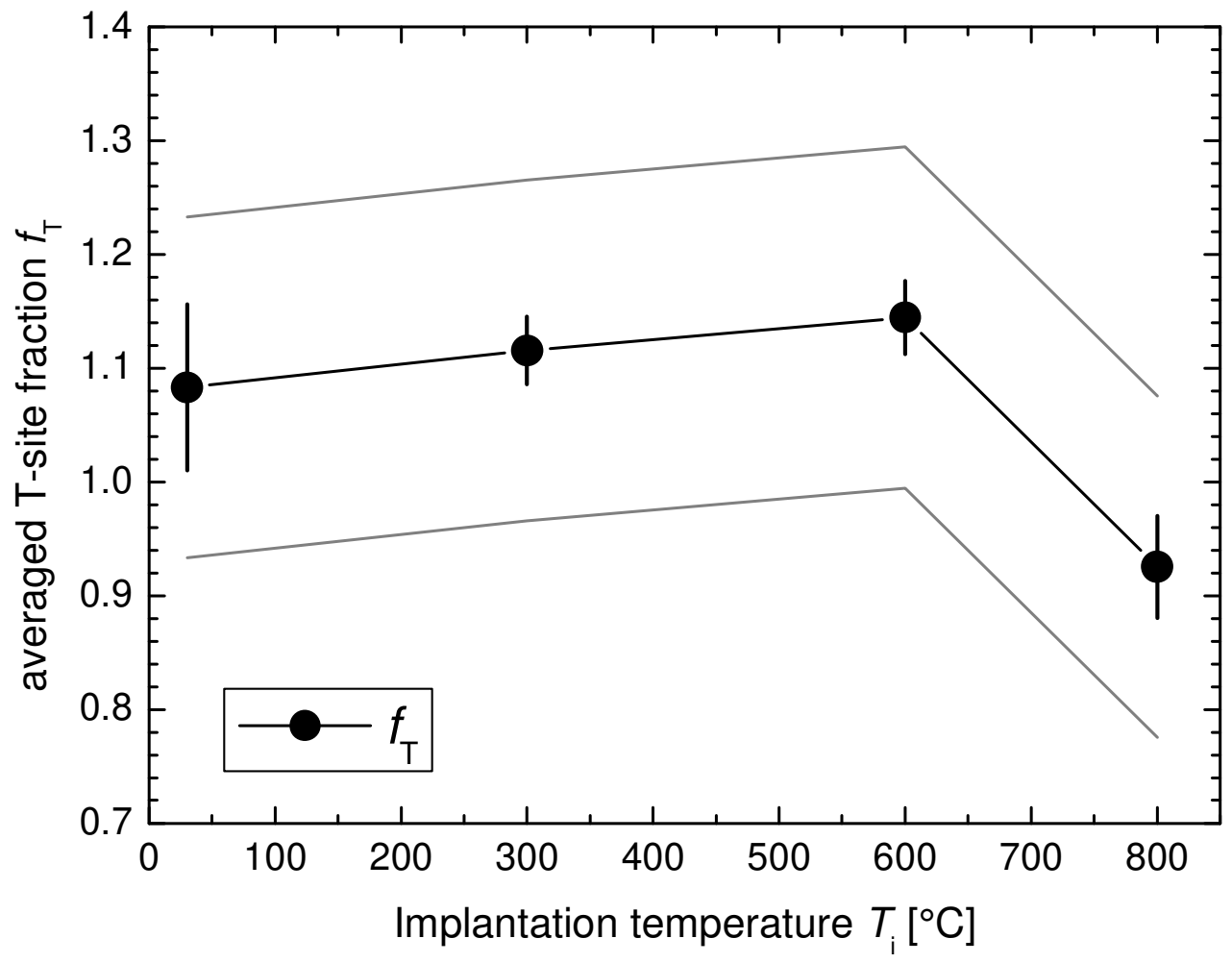

FIG 6. Fractions $f_T$ of $^6$He on T sites as function of implantation temperature $T_i$, obtained from averages of the fits to the <110>, <211>, <100>, and <111> patterns. The individual error bars represent the standard deviations derived from the separate measurements along the four different crystallographic directions, which we consider to be indicative of experimental *precision*. The absolute error of the fractions is estimated to be around ±15% as is indicated by the two extra lines, representing experimental *accuracy* due to possible errors in background correction. Such errors are also the likely cause of the derived fractions above 100%.

[36]. The dependence of the T-site fraction on the implantation temperature is plotted in Fig. 6, illustrating there is initially a slight upward trend from RT to 600°C, then at 800°C a decrease that reflects an ~18% drop of the anisotropy of the channeling patterns and which is consistent among the four measured crystallographic directions. This loss of anisotropy could not be identified as being the result of another type of high-symmetry lattice site being occupied (we mentioned already that two-site fits could not identify such a site, even at the implantation temperature of 800°C). However, what cannot be ruled out, is that some type of lattice site is involved which causes such low anisotropies in channeling effects, that they cannot be distinguished from so-called "random" sites, which are assumed to cause no anisotropies at all.

The most plausible explanation for the loss of angular anisotropy at 800°C, is the onset of interstitial migration of $^6$He, causing it to hop from one lattice site to another during its nuclear lifetime. Such migration processes can manifest themselves in emission channeling experiments most commonly via two phenomena:

A) a small number of atomic jumps that lead to the emitter atom occupying a different type of lattice site, i.e., essentially a short-range microscopic process. Prominent examples would be the lattice site changes of short-lived $^8$Li ($t_{1/2}$=838 ms) from T to substitutional cation sites in III-V and II-VI semiconductors [43,56-59]. These were interpreted as due to interstitial $^8$Li starting to migrate and combining with a cation vacancy during its lifetime, with associated Li migration energies $E_M$ from 0.32-1.7 eV (a more detailed overview is provided in the Supplemental Material [36]). The fact that these site changes were readily observable shows that the number of vacancies created by implantation of $^8$Li is sufficient to allow for the formation of substitutional $^8$Li on the timescale of roughly a second following its implantation. In the mentioned compound semiconductors this was likely aided by the fact that interstitial Li exists as positively charged ion $Li^+$ while the cation vacancies are negatively charged, resulting in attractive Coulomb interaction and the formation of stable substitutional Li. Interestingly, in the case of $^8$Li in Si [43] and Ge [59], although interstitial Li diffuses with migration energies of 0.655 eV and 0.512 eV, respectively, the capture of $^8$Li on substitutional sites was only observed in highly $n$-type Si, and in most other cases replaced by site changes to bond-center (BC) sites, indicating the formation of $^8$Li in split-vacancy configuration inside a double vacancy. Since the number of vacancies created during the implantation of $^6$He and its nuclear lifetime are not much different from the case of $^8$Li, one hence would expect that similar site changes to substitutional or BC sites should also be observable for $^6$He, if the resulting complexes are energetically favorable in comparison to simple interstitial He. The fact that this was not the case, indicates that substitutional He or He in split-vacancy configuration are not energetically very favorable configurations in diamond.

B) a large number of atomic jumps leading to a significant widening of the implantation profile, i.e., long-range diffusion where emitter atoms can reach the surface of the sample, where they are trapped or evaporated, and other emitter atoms reaching large depths inside the sample, where the channeling effects of the $\beta^-$ particles are destroyed due to the increased path length on the way out of the sample. Hence in case B, both diffusion to the surface and into the bulk lead to a loss of channeling effects, while the nature of the lattice site stays the same. Such examples have been found, e.g., for the cases of $^{107m}$Ag (44.3 s) and $^{109m}$Ag (39.8 s) in CdTe [60], where for a moderate increase in measurement temperature from RT to 100°C, the substitutional fraction of Ag dropped from 80-95% to 0%, which was accompanied by loss of more than 60% of the radioactivity from the sample due to out-diffusion. Another example is the dissociative out-diffusion of substitutional $^8$Li from InSb around 200°C, leading to complete loss of channeling anisotropy and a pronounced shift in its α-energy spectrum [43].

We can then use scenarios A and B to provide an estimate on the activation energy $E_M$ of interstitial migration of He, as follows. The diffusivity $D$ at temperature $T$ is given by the well-known Arrhenius relation

$$D = D_0 \exp\left[-\frac{E_M}{kT}\right] \quad (1)$$

with the entropy constant $D_0$ given by [43]

$$D_0 = \frac{l^2}{6}\nu_0 N_{NN} \quad (2)$$

where $l$=1.545 Å is the jump width from one interstitial T site to the next, $N_{NN}$=4 the number of nearest-neighbor T sites in the diamond structure, and the attempt frequency $\nu_0$ can be estimated as follows. The calculated effective attempt frequency of $^4$He in Ref. [33] is $\nu_0(^4He)$=29.3 THz, which, assuming an Einstein oscillator model, i.e., $\nu_0(^6He)=[4/6]^{1/2}\,\nu_0(^4He)$, gives $\nu_0(^6He)$=23.9 THz. This results in the entropy constants $D_0(^4He)=4.66\times10^{-3}$ cm$^2$s$^{-1}$ and $D_0(^6He)=3.80\times10^{-3}$ cm$^2$s$^{-1}$. The mean one-dimensionally projected diffusion width $R$ during a time period $\tau$(in our case $\tau$=1.164 s being the radioactive lifetime of $^6$He) is given by

$$R = \sqrt{2D\tau} \quad (3)$$

which means we get from eqs. (1)-(3) the relation

$$E_M = kT\ln\left[\frac{2D_0\tau}{R^2}\right] = kT\ln\left[\frac{l^2\nu_0 N_{NN}\tau}{3R^2}\right] \quad (4)$$

If we take $R$=1375 Å equal to the implantation depth (scenario B), this yields $E_M$=1.63 eV; if we take $R$=1.545 Å equal to the distance between adjacent T sites, hence one atomic jump only (limiting case scenario A), we get $E_M$=2.89 eV.

The only other chemical element for which, so far, the occupancy of tetrahedral interstitial T sites has been experimentally established in diamond, is Li [61], obtained from emission channeling experiments using the short-lived radioactive isotope $^8$Li (838 ms). It therefore seems as if only very small elements are able to occupy the interstitial T position in the tightly packed lattice of diamond: while He is the smallest atom found in nature, with an atomic radius of only 0.31 Å, Li is with 1.52 Å already much larger but might exist in +1 ionized form with ionic radius of 0.76 Å. No indication for diffusion of $^8$Li in diamond was observed for temperatures up to 630°C during its lifetime of $\tau$=1.209 s, pointing at an activation energy for migration $E_M$>1.25 eV [61].

As we have mentioned above, due to the short 807 ms half life of $^6$He there is no persistent accumulation of He in the sample, with the maximum concentration in the peak of the implantation profile (neglecting possible effects of diffusion) remaining at values around 1.1×10$^{14}$ atoms/cm$^3$. Hence it is quite unlikely that under our experimental conditions accumulation of He, e.g., into bubbles, would occur.

## IV. CONCLUSIONS

Without doubt, the most frequently found lattice site of ion implanted He in diamond is the tetrahedral interstitial T site. The possible fractions of He on other lattice sites of higher symmetry must be small for all implantation temperatures. In particular, none of the sites theoretically predicted to be associated with He color centers in diamond [35] could be identified. However, implantation at 800°C led to partial loss of channeling anisotropy, which we suggest to be the result of a fraction of ~18% of $^6$He changing its site to a position of very low crystal symmetry or due to long-range diffusion of $^6$He during its lifetime. Assuming these two limiting cases, we estimate the activation energy of interstitial migration between 1.63-2.89 eV, which is rather well in accordance with theoretical predictions of 1.41-2.36 eV [31-33,35]. Activation energies around 2 eV for interstitial He migration would mean that *simple interstitial* He cannot be stable in diamond on geological time scales. Thus, for He to remain inside the diamond, it should be bound to some defect in the material or exist in another form such as within inclusions of other minerals or liquids, or possibly small He bubbles.

## ACKNOWLEDGMENTS

We appreciate the support of the ISOLDE Collaboration and technical teams. This work was funded by the Portuguese Foundation for Science and Technology (Fundação para a Ciência e a Tecnologia FCT, projects 2024.00223.CERN DOI 10.54499/2024.00223.CERN , UID/04349/2025 DOI 10.54499/UID/04349/2025), by the Research Foundation Flanders (FWO, Belgium), and by the KU Leuven. The EU Horizon Europe Framework supported ISOLDE beam times through Grant Agreement 101057511 (EURO-LABS). We thank Z. Santha, G. Thiering, and A. Gali for helpful discussions and making available their calculated He configurations.

* Corresponding author.
uwahl@ctn.tecnico.ulisboa.pt

**Supplemental material to:**

# Lattice location of implanted He in diamond

U. Wahl, J. G. Correia, A. Costa, B. Biesmans, G. Magchiels, S. M. Tunhuma, A. Lamelas, A. Vantomme, and L. M. C. Pereira

**Note: numbers of references are with respect to those in the main paper**

## 1. Literature overview of theoretical studies on He in diamond

Reference [30], using a tight-binding potential approach but not addressing the question of stability of different He configurations in the diamond lattice, concluded that interstitial $He_i$ should be symmetrically positioned on tetrahedral T sites, i.e., without any displacements from these interstitial positions of highest crystallographic symmetry, while substitutional $He_s$ was predicted to undergo a displacement of about 0.3 Å from the ideal S site.

Reference [31] predicted that $He_i$ on T sites has the lowest formation enthalpy $\Delta H$ with $\Delta H(He_T)$=6.3 eV, vs. $\Delta H(He_s)$=10 eV for substitutional $He_s$, and $\Delta H(HeV)$=12 eV for the He*V* complex, the latter exhibiting a split-vacancy configuration, i.e., the He atom occupying the center position within a double-vacancy (identical to a bond-center position BC). The binding energy of He*V* was calculated as 2.2 eV, from which the activation energy for dissociation into interstitial $He_i$ and *V*, approximated by the sum of the binding and diffusion energies, was estimated as 4.5 eV. The hexagonal H site was projected to be the saddle point of the diffusion path of interstitial He, i.e., a T-H-T migration path, for which the activation energy $E_M$ was predicted as $E_M$=2.3 eV. Finally, the local mode frequency $\nu_0$ of $^4$He vibration on the T site was calculated as $\nu_0(^4He)$=14 THz.

An interesting study based on atomistic ab-initio and tight-binding models [32] predicted the activation energy for interstitial migration of He as $E_M$=1.41 eV at ambient pressure, which was expected to lower to $E_M$=1.20 eV at a pressure of 20 GPa, characteristic for diamond-bearing layers inside the earth mantle. Besides interstitial $He_T$ on T sites and substitutional sites, which were assessed to have the configurations already mentioned above with respect to Ref. [30], this work also addressed the structure and stability of possible small He inclusions in diamond, i.e., several He atoms inside multi-vacancy complexes. It was found that "a single carbon vacancy can accommodate up to 4 helium atoms in an energetically favorable position compared to the interstitial defect", and with respect to multiple vacancies that "the most stable site of helium in diamond is in the cavities".

Reference [33] also forecast the T site to be the most stable position of He in diamond, with an activation energy of $E_M$=1.97 eV for T-H-T migration, and local mode frequencies of $\nu_0(^4He)$=29.3 THz and $\nu_0(^3He)$=33.8 THz.

The atomic structures of noble gases He, Ne, Ar, Kr and Xe in diamond were systematically investigated in Ref. [34], considering substitutional S and interstitial T sites, as well as split-vacancy configurations as possible defect geometries. $He_T$ was predicted on ideal T sites with only an outward relaxation of the four surrounding C atoms, while $He_s$ should sit off-center from the ideal S position, around 0.5 Å towards AB sites, and He within He*V* in the ideal BC position. The predicted formation enthalpies were $\Delta H(He_T)$=6.22 eV, $\Delta H(He_s)$=18.58 eV, and $\Delta H(HeV)$=30.30 eV, i.e., tetrahedral interstitial $He_T$ being the most stable configuration. No activation energies for diffusion were addressed in this reference.

Very recent computational results [35] also predict that interstitial He atoms reside in the T site and only slightly distort the lattice around them in a symmetric breathing mode distortion. The barrier energy for the migration of the $He_T$ defect was calculated as 2.36 eV. Besides interstitial He, also substitutional He was considered, however, here it was found that He on substitutional sites should undergo large lattice relaxations that move it a considerable distance away from the ideal substitutional position, 0.695 Å along the <100> direction ($C_{2v}$ symmetry), which would be the SP→C 0.60 site with our nomenclature, see Fig. 1 (f), or 0.794 Å along the <111> direction ($C_{3v}$ symmetry) which is very close to the AB site, see Fig. 1 (e).

## 2. Literature overview of experimental results on lattice sites of He in diamond and the related semiconductors Si and 6H-SiC

Experimental results on the lattice location of He in semiconductors are scarce but there exists some limited work on diamond and its close relatives Si and 6H-SiC. Allen used a combination of Rutherford Backscattering Spectrometry and $^3$He(d,p)$^4$He Nuclear Reaction Analysis in channeling mode (RBS/C, NRA/C) to investigate the lattice location of ~70 keV implanted $^3$He in natural diamond and Si [37], as well as 6H-SiC [38]. The implanted fluences were $4.2\times10^{15}$ cm$^{-2}$ for diamond, $5\times10^{15}$ cm$^{-2}$ for Si, and $3.7\times10^{15}$ cm$^{-2}$ for 6H-SiC. The results indicated for diamond and Si qualitative preferences of He for interstitial T sites. In the case of 6H-SiC, the chosen angular scans, which were the <0001> axis and the (1−210) plane, while suitable to rule out a major occupation of the so-called octahedral interstitial O sites of the 6H structure, could not be used in order to distinguish between substitutional Si, C or tetrahedral interstitial T sites. The studies of Allen thus gave a first idea what might be the lattice site preference of He in these semiconductors. However, due to the fact that only two to three one-dimensional angular scans were measured in each case and the experimental results not quantitatively analyzed by means of comparison to theoretically predicted channeling and blocking yields for different lattice sites, the results were not very conclusive regarding the exact lattice location(s) of He and the fraction of implanted atoms which occupy it. For diamond, we estimate from the observation of dips in the yield of the $^{3}He(d,p)^{4}He$ reaction in Ref. [36] for deuteron incidence along <111> and (110), which were of similar angular width as the signals from backscattered deuterons, that a substantial part of $^{3}He$ was quite well aligned with this axis and this plane. On the other hand, a channeling peak for deuteron incidence along the <110> direction points to interstitial T sites. From the minimum yield of the <111> effect one can estimate ~33% of T sites, the rest on random sites.

### 3. Comments on literature overview of He diffusion data in diamond

The activation energies $E_A$ and entropy constants $D_0$ derived from experimental observations depend on a multitude of experimental conditions, like type of sample (natural vs artificially grown by CVD = Chemical Vapor Deposition, powder or polycrystalline vs single crystal), how He was introduced (naturally occurring, or ion implantation), which isotope was looked at (e.g., cosmogenic $^{3}He$ vs $^{4}He$ resulting from alpha decay) which method was performed to detect it (e.g., mass spectroscopy, meaning one measures when He is released from the sample, vs NRA = Nuclear Reaction Analysis, which can probe changes in the He depth profile but also its loss), what temperatures and how many temperature steps were used in the study, which model assumptions are assumed when deriving the values from experimental data etc. Since, a priori, the type of diffusion mechanism that occurs in the experiment is unknown, one can compare to theoretical modelling (and also data for other diffusing elements or materials that are better understood) in order to get insight into this. The spread in experimental activation energies and entropy constants hints at several different diffusion or release mechanisms being at work for He in diamond, and the difference that most values show towards theoretically predicted values for simple interstitial migration, suggest that in many cases diffusion or release was not due to simple interstitial migration. This is in particular suggested if substantial He release was only observed at temperatures close to where the material already completely disintegrates (1700-2000°C), which seems to have been the case for the three mass spectroscopic studies of He in natural diamond. In this case.

### 4. Some details on "many-beam" simulations for $\beta^-$ particles from $^{6}He$ in the diamond structure

Details on how the many-beam simulations of electron channeling in diamond were performed can be found in the supplementary materials of Refs. [53,54]. Compared to these descriptions, the many-beam simulations used here differ in two details. First, angular patterns were calculated in $x$- and $y$-direction from −3° to +3° in steps of 0.025°, i.e., in two times finer mesh. Second, to account for the $\beta^-$ spectrum of $^{6}He$, many-beam simulations were performed for a number of discrete electron energies up to its endpoint energy of 3506 keV (step widths of 25 keV in the energy range 25-800 keV, 50 keV in the range 800-1400 keV, 100 keV in the range 1400-2500 keV, 250 keV in the range 2500-3500 keV, and weighting as indicated by the histogram according to the $\beta^-$ spectrum of $^{6}He$ (Fig. S1).

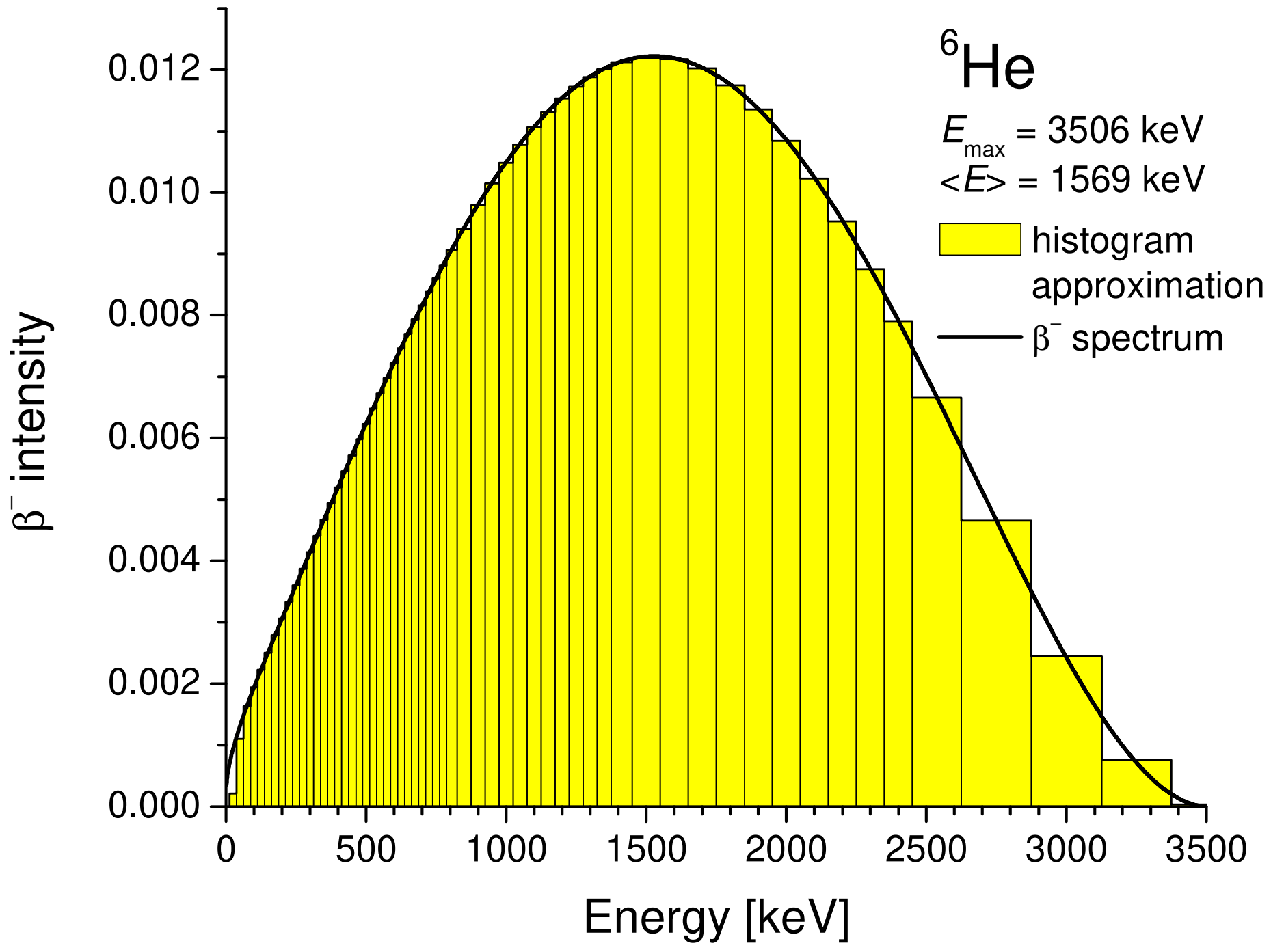

**FIG. S1.** $\beta^-$ spectrum of $^{6}He$ and the histogram of the energy values used to approximate it in the "many-beam" calculations. The intensity of the first two histogram bars is lower than in the $\beta^-$ spectrum due to the detector threshold around 35 keV.

## 5. Counting statistics of emission channeling patterns

Counting statistics of all $\beta^-$ emission channeling pattern measurements used in this manuscript. The first column indicates the implantation and measurement temperature. The number of events (counts) is given as total number for the whole detector, as well as per pixel (obtained by dividing total # events by 484).

| | <100> total # events | <100> per pixel | <111> total # events | <111> per pixel | <211> total # events | <211> per pixel | <110> total # events | <110> per pixel |
|---|---|---|---|---|---|---|---|---|
| $T_i$=RT=30°C | 181744 | 375 | 1472410 | 3042 | 541184 | 1118 | 2975470 | 6148 |
| $T_i$=RT=30°C | 1113080 | 2300 | | | | | | |
| $T_i$=RT=30°C | 1845019 | 3812 | | | | | | |
| $T_i$=300°C | 1001580 | 2069 | 1502481 | 3104 | 2992299 | 6182 | 2099598 | 4338 |
| $T_i$=600°C | 1003731 | 2074 | 2928386 | 6050 | 3004952 | 6209 | 3020443 | 6240 |
| $T_i$=800°C | 873542 | 1804 | 3005094 | 6209 | 3057952 | 6318 | 3324316 | 6868 |

**6. Channeling patterns for 600°C implantation**

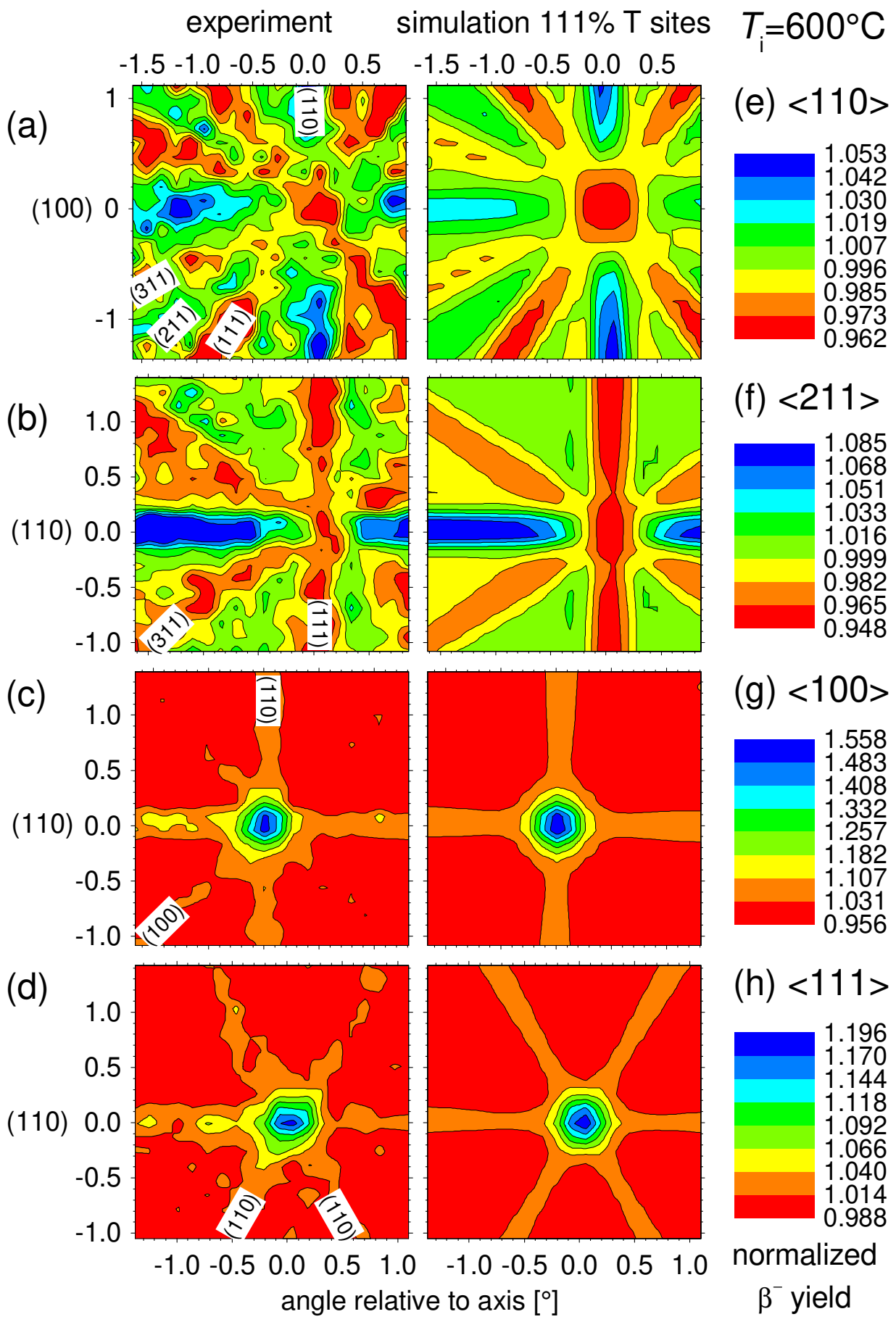

**FIG. S2.** (a)-(d) Angular distributions of $\beta^-$ emission yields from $^6$He in diamond, during implantation at 600°C, around the crystallographic directions <110>, <211>, <100>, and <111>. (e)-(h) are the best fits of simulated patterns, corresponding to 111% of $^6$He on tetrahedral interstitial (T) sites.

## 7. Chi square analysis of possible $^{6}$He lattice sites

### 7.1 One-site fits

**FIG. S3.** Relative chi square of ***one-site*** fits as a function of the position along the <111> direction (a)-(d), <100> direction (e)-(h), and <110> direction (i)-(l). The zero point of the distance scale has been fixed at the T site for all displacements. Relative chi squares of fit are given as $\chi^2/\chi^2(\mathrm{T})$, i.e., obtained by dividing by the chi square $\chi^2(\mathrm{T})$ resulting from T site fits. Note that <111> and <100> results cannot be used to distinguish between all possible sites since, due to the crystal symmetry, e.g., their channeling patterns for S and T or BC, AB and H sites are equal. The three black curves in panels (b), (f), and (j) all represent <100> measurements obtained at RT but with slightly different orientations towards the detector and different number of events/pattern. The plots have been scaled so as to provide good resolution near the $\chi^2$ minima and therefore values that fall above the chosen $\chi^2$ range are not shown in all cases. The curves shown in light gray colors correspond to fits with a negative site fraction ($f$<0), hence physically irrelevant results. Note that panels (a)-(d) are identical to Fig. 5 of the main manuscript and are included here for ease of comparison.

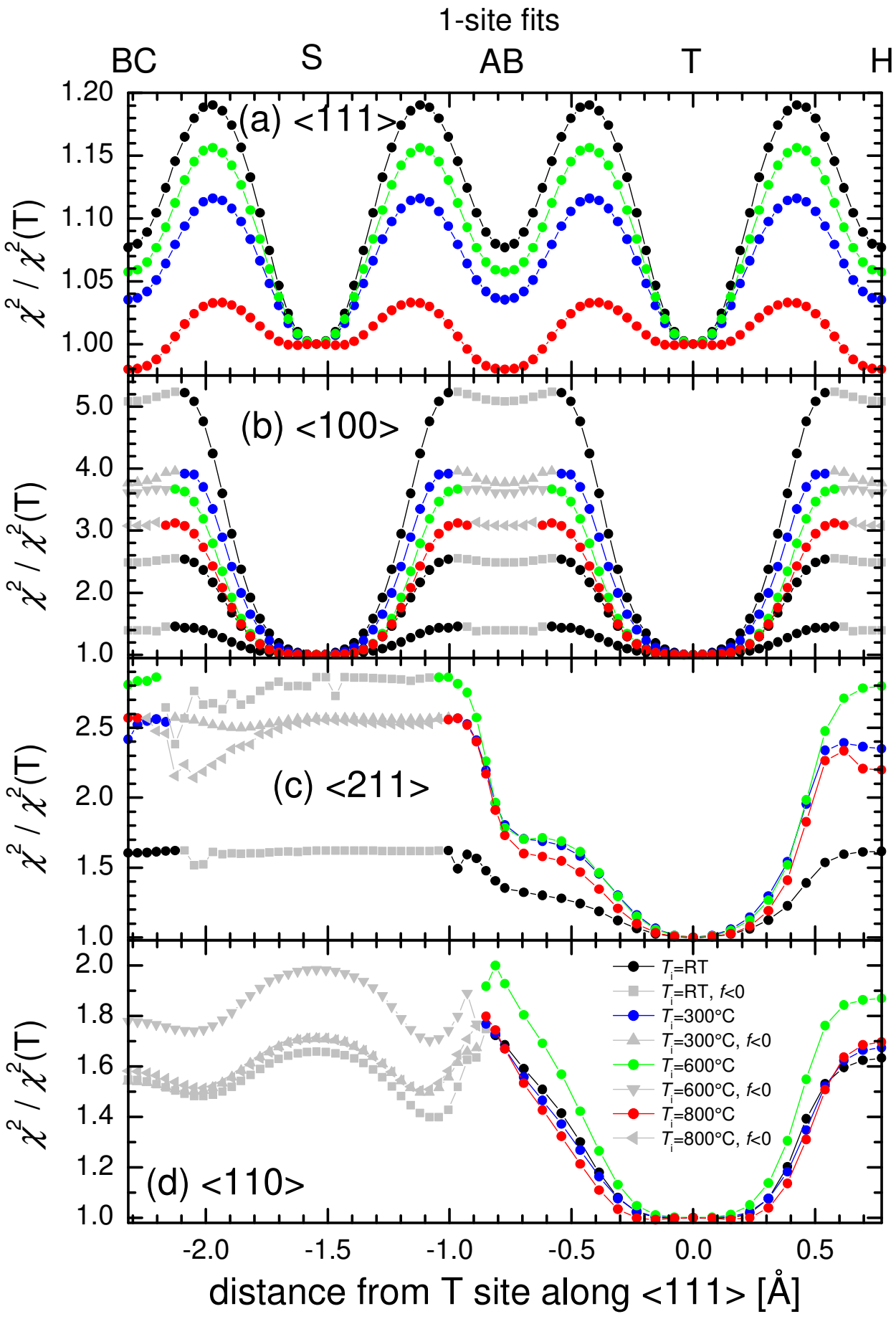

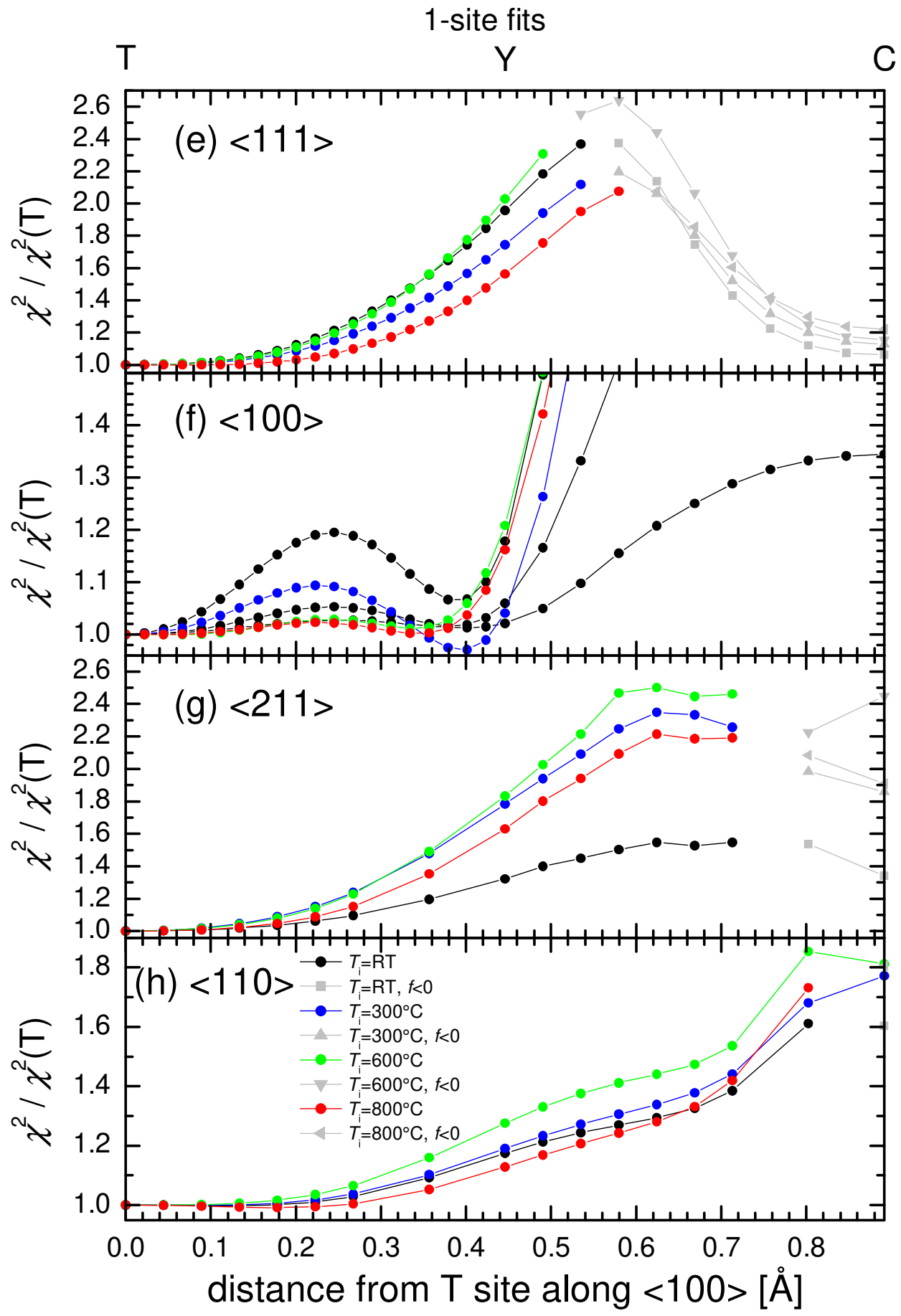

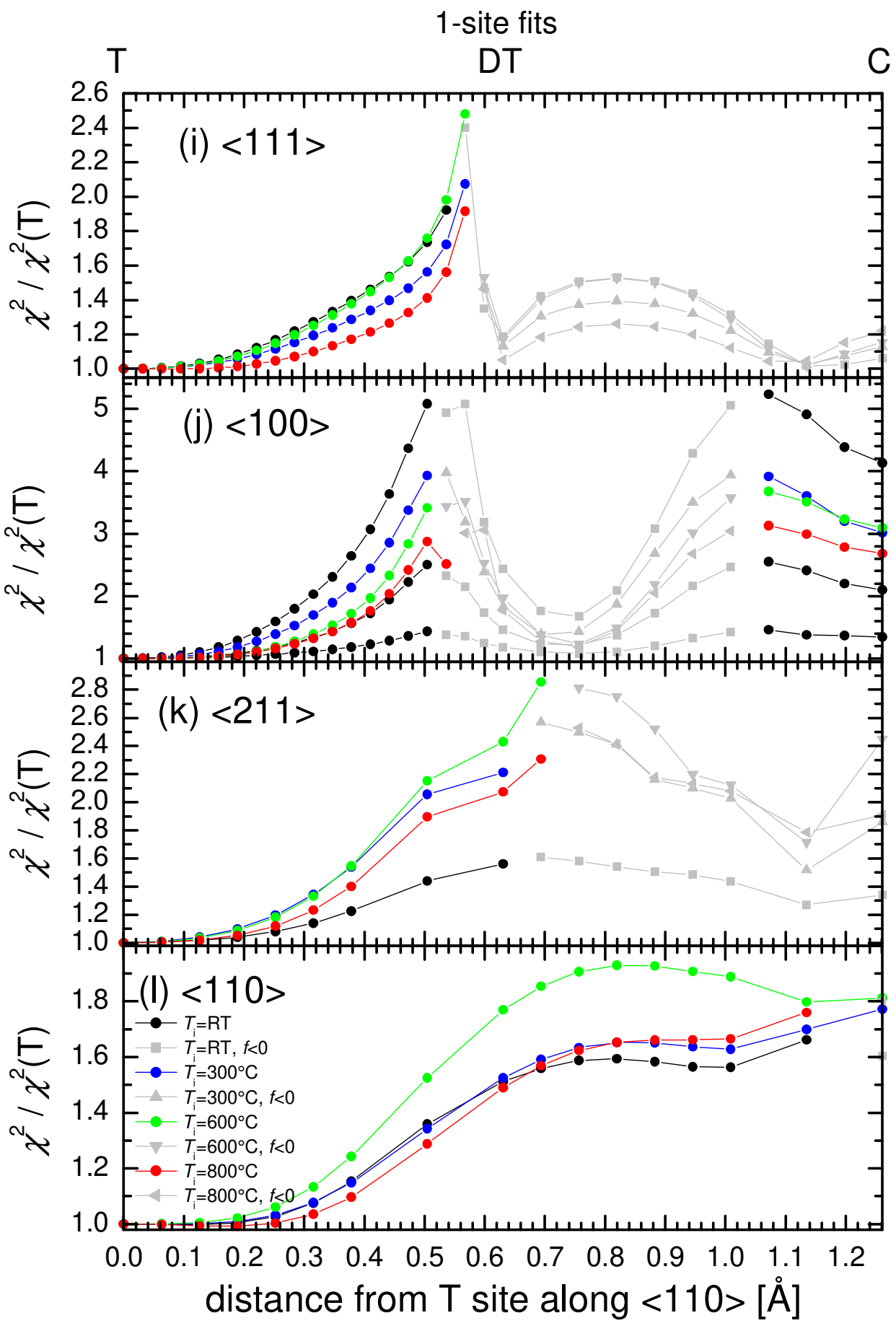

### 7.2 Two-site fits

**FIG. S4.** Relative chi square of ***two-site*** fits as a function of the position of the 2nd site along the <111> direction (a)-(d), <100> direction (e)-(h), and <110> direction (i)-(l), while the position of the 1st site is kept fixed at the T site . In contrast to Fig. S3, the zero point of the distance scale has been fixed at the S site for all displacements. Relative chi squares of fit are given as $\chi^2/\chi^2(\mathrm{T})$, i.e. obtained by dividing by the chi square $\chi^2(\mathrm{T})$ resulting from T site fits only. Note that <111> and <100> results cannot be used to distinguish between all possible sites since, due to the crystal symmetry, their channeling patterns are identical for a number of sites, e.g. for S and T or BC, AB and H sites. The three black curves in panels (b), (f), and (j) all represent <100> measurements obtained at room temperature but with slightly different orientations towards the detector and different number of events/pattern. The curves shown in light gray colors correspond to fits with at least one negative site fraction ($f_1$<0 or $f_2$<0), hence physically irrelevant results.

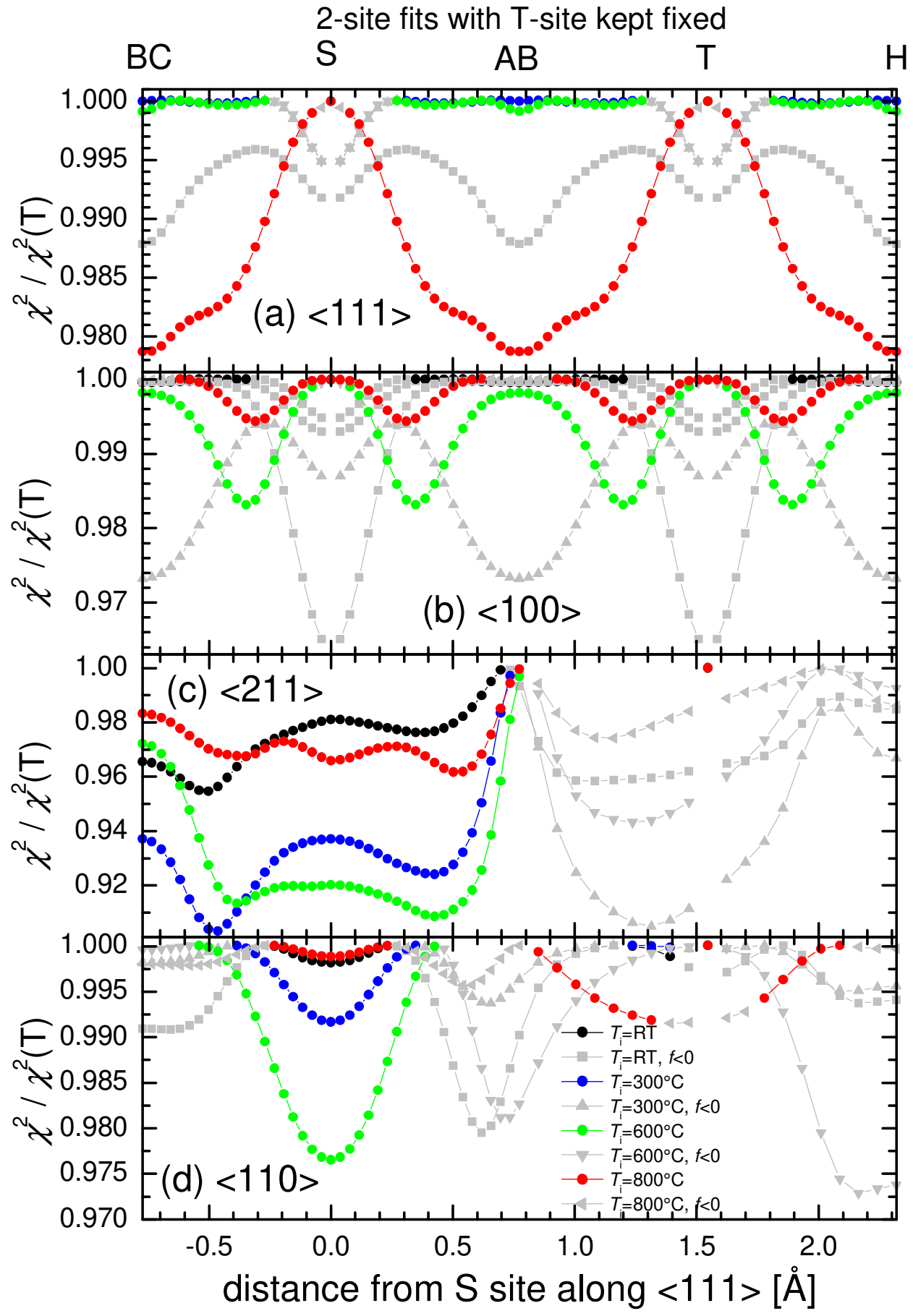

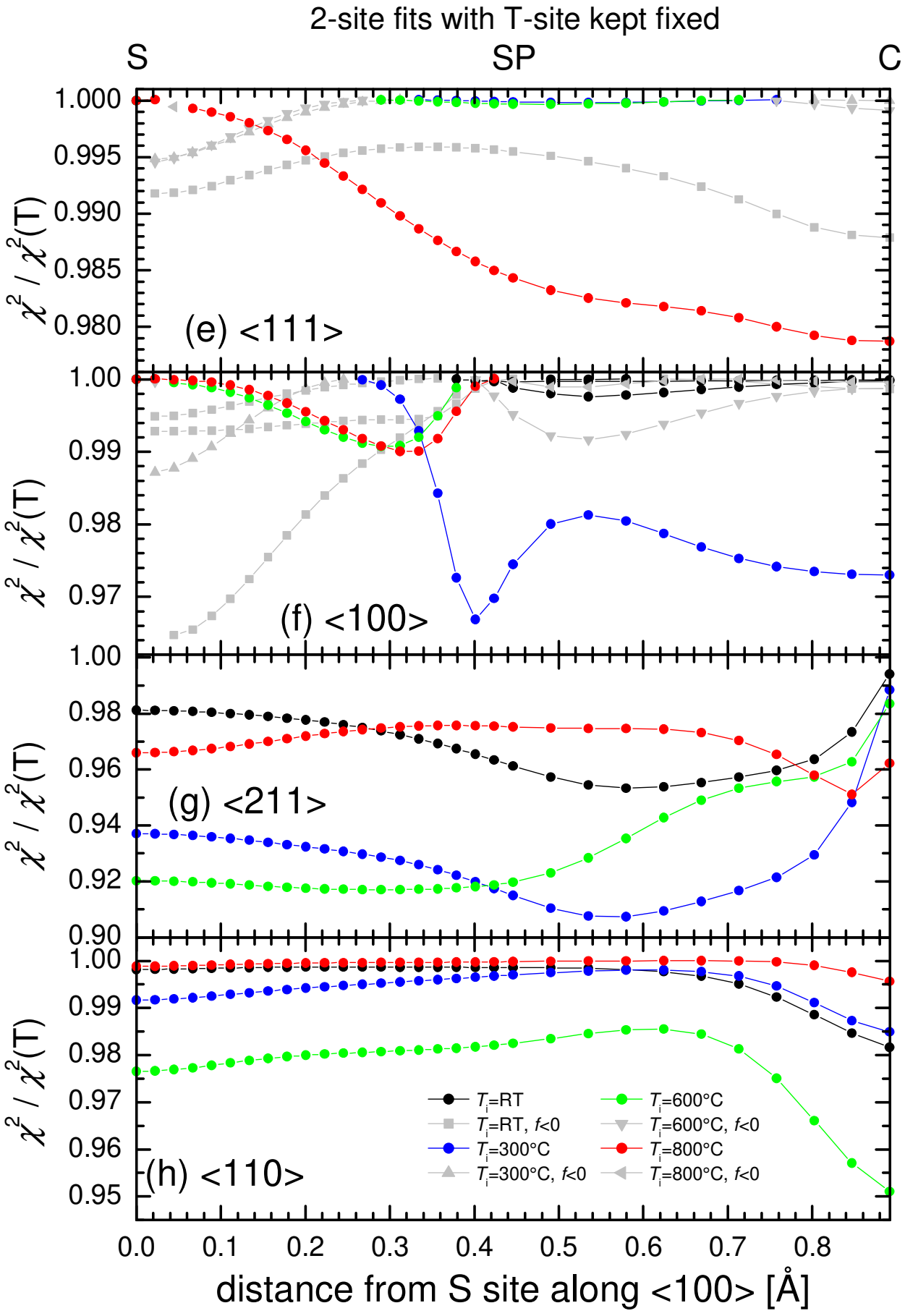

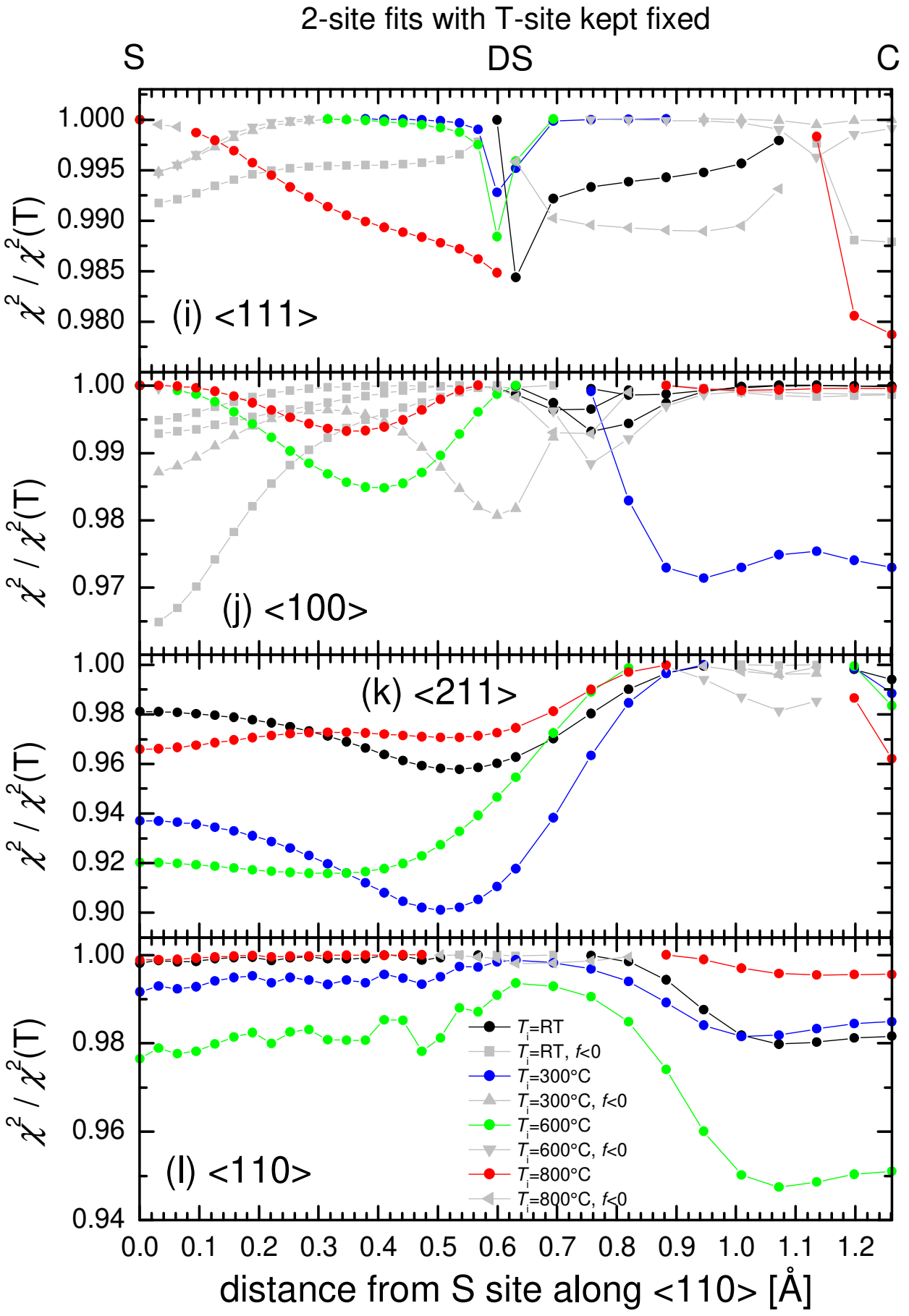

## 8. Comparison of experimental interstitial migration energies of $^{6}He$ and $^{8}Li$ in semiconductors

Overview on the lattice site changes of short-lived $^{8}Li$ ($t_{1/2}$=838 ms) from T to substitutional cation sites in elemental, III-V and II-VI semiconductors [42,43,56-59]. In the compound semiconductors, about 60-70% of the implanted $^{8}Li$ was found on tetrahedral interstitial sites in the as-implanted state at low temperatures, with the remainder on “random” sites, i.e. on sites of very low crystal symmetry. Upon implantation at elevated temperatures, practically all of the tetrahedral interstitial $^{8}Li$ was found on substitutional sites, which was interpreted as due to interstitial $^{8}Li$ starting to migrate and combining with a cation vacancy during its lifetime. The temperature ranges where the site changes T→S occur, vary considerably, being as low 155-180 K in the case of InSb, but as high as 377-477°C=650-750 K in the case of GaN and AlN.

| Semiconductor | Structure | Site change process | $E_M$($^{8}$Li) [eV], Ref. | $E_M$($^{6}$He) [eV] |
|---|---|---|---|---|
| C (diamond) | diamond | T → T | >1.25 [61] | 1.63-2.89 [this work] |
| Si | diamond | T → BC, T → S | 0.655 [43] | |
| Ge | diamond | T → BC | 0.512 [59] | |
| AlN | wurtzite | T → $S_{Al}$ | ~1.7 [56] | |
| GaN | wurtzite | T → $S_{Ga}$ | ~1.7 [56] | |
| GaAs | cubic zincblende | T → $S_{Ga}$ | 0.58-0.60 [43] | |
| GaP | cubic zincblende | T → $S_{Ga}$ | 1.14-1.19 [43] | |
| InP | cubic zincblende | T → $S_{In}$ | 0.87-0.91 [43] | |
| InSb | cubic zincblende | T → $S_{In}$ | 0.37-0.39 [43] | |
| CdTe | cubic zincblende | T → $S_{Cd}$ | 0.33-0.37 [57] | |
| ZnSe | cubic zincblende | T → $S_{Zn}$ | 0.45-0.55 [58] | |
| ZnTe | cubic zincblende | T → $S_{Zn}$ | 0.32-0.42 [58] | |

## 9. Background correction

Fast electrons moving in matter are subject to pronounced scattering; it is therefore not possible to measure electron emission channeling effects that are not accompanied by a background of scattered electrons. The major contributions are electrons that are backscattered from inside the sample or from the walls of the vacuum chamber. In addition, there exists a background of gammas that are either emitted by the sample itself or originate from sources in the neighborhood. All types of background contribute to the measured patterns, usually with a rather homogeneous count rate, and thus lower the anisotropy, which means the background increases the so-called random fraction $f_R$ obtained during a fit. In order to correct for this, the fitted fractions need to be multiplied by a correction factor, as is outlined, e.g., in Ref. [47]. Note that this procedure is mathematically equivalent to subtracting a flat baseline from the measured patterns.

The contribution of backscattered electrons was estimated by performing GEANT4 Monte Carlo simulations of electron trajectories, taking into account the main geometrical features and elemental composition of the sample (diamond), the sample holder (Mo), and the vacuum chamber (stainless steel), as well as the energy distribution of $\beta^-$ particles emitted by $^{6}He$ (Fig. S1). According to the simulations, electrons that reach the detector but were initially emitted with a velocity vector pointing away from it (29.2%), as well as electrons that reached the detector after moving inside the walls of the vacuum chamber (26.1%), resulted in an overall background of 46.9% (this percentage is somewhat smaller than the sum of 29.2% + 26.1% = 55.3% since in the simulations some of the electrons are backscattered both from inside the sample and by the walls), meaning the site fractions $f_T$ obtained in the fits were subject to an electron scattering correction factor of 1.88.

Background caused by gamma radiation emitted by the radioactive sample itself, as well as from radioactive sources in the vicinity and cosmic radiation, can be experimentally estimated by closing a shutter valve in front of the detector, which stops all electrons from the sample but allows most of the $\gamma$ particles to pass. During on-line experiments inside the ISOLDE hall, the background from cosmic radiation and other natural sources is negligible since the $\beta^-$ count rate resulting from the sample is much higher than these contributions. While the isotope $^6$He itself emits no gammas, there remains the contribution from gammas due to radioactive isotopes in the vicinity, mostly those resulting from the ISOLDE production target which is activated by the proton beam. Since $^6$He is a comparatively weak beam provided by ISOLDE, the target was run with maximum proton beam intensity of 2 µA available, resulting in a rather high detector base count rate of ~183 Hz, while the count rate was ~900 Hz with the sample exposed to $^6$He. Therefore, a gamma correction factor of 1.25 was estimated.

The fitted fractions were thus multiplied by an overall background correction factor of 2.35. The absolute error in performing the background correction as outlined above is estimated at ~±10−15%. The fact that the fitted fractions after background correction reached values above 100% indicates that the background was overestimated, but still within the expected range.

In addition to relatively high background, $^6$He experiments, due to its very light mass, are also subject to comparatively large implantation depths in the diamond sample, being up to 10 times deeper than in typical other cases. For instance, the projected range and straggling of 30 keV implantations into diamond is 1375(299) Å for $^6$He, much larger than 291(85) Å for $^{27}$Mg, 149(36) Å for $^{75}$Ge, or 132(24) Å for $^{121}$Sn [52]. The increased depth means that electrons channeling inside the sample face a much greater path length on their way out, so that dechanneling due to scattering with diamond electrons or thermally moving C atoms is rather more relevant than in the case of other isotopes, reducing the maximum observable channeling effects to larger extent. While the effect is taken into account during the simulation of channeling patterns using the many-beam approach via first-order perturbation theory, the approach is approximate and in case of $^6$He experiments due to the increased path length may be subject to larger errors than for other isotopes. However, possible such errors are rather temperature-independent, unless there should occur a major change in the depth profile, e.g., due to diffusion.

## 10. Error estimates on fractions: precision *vs* accuracy

It is important to point out the difference between ***accuracy*** and ***precision*** in our assessment of site fractions: ***Accuracy*** refers to how close a measurement is to the true or accepted value. ***Precision*** refers to how close repeated measurements are to each other, regardless of whether they are correct, hence *precision* refers to reproducibility. In that respect, the ***accuracy*** of our fractions is not overly good, since we estimate the relative error of the sum of all assigned fractions to be 10-15% off the real value, mostly a consequence of the overall background correction factor of 2.35±(10-15%), that is common to all measurements. There are other sources of systematic errors but the background correction is the most relevant. Others are, e.g., insufficient knowledge of the depth profile of implanted atoms, and related to that, theoretical assumptions how the electrons dechannel in the crystal, insufficient knowledge of the angular resolution, which depends on size and shape of the beam spot on the sample, and on the position resolution of the detector, which varies with the $\beta^-$ energy.

On the other hand, the ***precision*** of repeated measurements can be very high: e.g. when repeating the measurement of a crystallographic direction at a different temperature but otherwise unchanged conditions, the error in the fitted fractions is basically determined only by counting statistics and the resulting error in the fraction fit parameter; one can reach 0.1%, which is the case for the measurements with highest statistics in this manuscript (~6 million events/pattern, ~6200 counts/pixel). Even for an initial measurement with ~17× lower statistics (181744 events/pattern, 375 counts/pixel), the fit error was only 0.3%. The reproducibility is less if the crystal has been rotated and then moved back to the old orientation (as was done for the measurements in this study in order to sequentially measure all 4 crystallographic directions at each temperature): because of finite goniometer precision (0.05° for the main goniometer rotation axis), one does not return to exactly the same angular position. Experience shows that this finite precision in angular position introduces errors in the reproducibility of repeated fraction measurements that are around the 1% level. Another option to assess the *precision* of equivalent measurements that should give the same fitted fractions, is to look at the divergence of fitted site fractions for the 4 different crystallographic directions. This is the approach we have taken in Fig. 6 of the main paper: the ***precision*** is indicated by the individual error bars that are calculated as the standard deviation of the fitted T-site fractions for the four different crystallographic directions. Then ±15% boundaries are indicated which reflect the finite ***accuracy*** as a consequence of possible errors in background correction and other systematic errors. A change in background correction factor would mean the whole curve would move up or down, not individual data points. Within the *precision* of the measurements, the relative ~18% decrease at 800°C is therefore certainly statistically relevant.